\documentclass{v19cosmo}

\usepackage{fontenc}
\usepackage{mathtools,amssymb,amsmath}
\usepackage{hyperref}

\newcommand{\Photo}{}

\begin{document}
\vspace*{4cm}
%\title{A Precise Dictionary Between Primordial Power Spectra and Effective Field Theory of Inflation Parameters}
\title{A Map Between Primordial Power Spectra and the Effective Field Theory of Inflation}

\author{Amel Durakovic \footnote{\href{mailto:amel@nbi.dk}{amel@nbi.dk}}\\[0.2cm]}

\address{Niels Bohr International Academy and Discovery Center, \\ Niels Bohr Institute, \\ Blegdamsvej 17, 2100 Copenhagen, Denmark \\[0.4cm] Technical University of Denmark \\ Fysikvej Building 309, \\ Kongens Lyngby, DK 2800, Denmark}

%\maketitle\abstracts{We develop a dictionary between changes in the primordial power spectrum and the changes in the parameters of the effective theory of inflation that induce them. The effective field theory of inflation contains only the curvature perturbation with two parameters: the slow-roll parameter $\epsilon$ as expansion parameter and the speed of sound $c_s$.}

\maketitle
\abstracts{We have developed a precise dictionary between the spectrum of primordial density fluctuations and the parameters of the effective field theory (EFT) of inflation that determine the primordial power spectrum (PPS). At lowest order the EFT contains two parameters: the slow-roll parameter $\epsilon$, which acts as an order parameter, and the speed of sound $c_s$. Applying second-order perturbation theory, we provide maps from the PPS to the EFT parameters that are precise up to the cube of the fractional change in the PPS $(\Delta \mathcal{P}/\mathcal{P})^3$, or less than $1\%$ for spectral features that modulate the PPS by $20\%$. While such features are not required when the underlying cosmological model is assumed to be $\Lambda$CDM they are necessary for alternative models that have no cosmological constant/dark energy. We verify the dictionary numerically and find those excursions in the slow-roll parameter that reproduce the PPS needed to fit Planck data for both $\Lambda$ and no-$\Lambda$ cosmological models.} %The reconstructions suggest that the no-$\Lambda$ cosmological model requires the slow-roll parameter to have fine structure in order to fit Planck data.} %The reconstructed slow-roll parameter from Planck data when the no-$\Lambda$ cosmology is assumed.} %As a corollary we can also reconstruct the scalar field potential that gives the desired PPS modulation.}

%The primordial power spectrum describes the variance of the Fourier coefficients of the curvature perturbations and the simplest models of inflation predict that it should be without characteristic scales, a power law that only mildly depends on scale. In more complicated scenarios, 

\section{Introduction}
Before reporting on our work \cite{Durakovic:2019kqq} which provides a map between changes in the parameters of an effective description of inflation and the modulations in the primordial fluctuations, and vice versa, we will recall the basics of inflation, the observational consequences and motivate the departure from the simplest realisation of inflation.

Inflation is a near-exponential expansion of space that results in $50$-$60$ $e$-folds of expansion, a $50$-$60$-fold multiplication of the scale factor by $e$. This large increase of the scale factor dilutes any exotic relics, flattens the space and stretches regions previously in causal contact beyond the horizon (the largest observable scales are only reentering today), solving the monopole, flatness and horizon problem, respectively.

Inflation is typically realised by a scalar field slowly rolling down an almost-flat non-zero potential. The low kinetic energy and non-zero potential act as an energy density with negative pressure supporting the near-exponential expansion. %, revealing today a uniformity on cosmological scales due to their first contact prior to inflation.

Inflation occurs when the slow-roll parameter $\epsilon = -\dot{H}/H^2 < 1$, and space-time is almost de Sitter when $\epsilon \ll 1$. This requires that the scalar field kinetic energy $\dot{\phi}^2/2$ is small as it also holds that $\epsilon = -\dot{\phi}^2/ (2H^2)$. In addition, for inflation to last long enough the change in the kinetic energy should be small motivating the introduction of a second parameter $\eta = -\ddot{\phi}/(H \dot{\phi}) \ll 1$.

As the scalar field $\phi$ is subject to quantum fluctuations there are accompanying fluctuations in energy-momentum which source curvature perturbations, fluctuations in the gravitational potential that lead to later structure formation.
%These are $\epsilon = -\dot{H}/H^2$ and $\eta = -\ddot{\phi}/(H \dot{\phi})$ which ensure that inflation occurs, requiring $\epsilon \ll 1$ and that it lasts long enough $\eta \ll 1$. In the case of a single field driving inflation $\epsilon = -\dot{\phi}^2/ (2H^2)$ and under slow roll $\epsilon \approx - M_\mathrm{Pl}^2/2 V'(\phi)/V(\phi)$ and $ \eta = M_{\mathrm{Pl}}^2 V''(\phi)/V$.
Inflation therefore also provides a quantitative description of primordial fluctuations as they are summarised in the primordial power spectrum (PPS). The PPS $P(k)$ is the variance of the Fourier coefficient $\mathcal{R}_\mathbf{k}$ of the curvature perturbation $\mathcal{R}$
\begin{align}
\langle \mathcal{R}_{\mathbf{k}} \mathcal{R}_{\mathbf{k'}} \rangle = (2 \pi)^3 \delta^{(3)}(\mathbf{k}+\mathbf{k'}) P(k)
\end{align}
which can be scaled by a power of the wave numbers to give the \emph{dimensionless} PPS $\mathcal{P}(k) \equiv k^3 P(k)/(2 \pi^2)$, henceforth only referred to as the PPS. The curvature perturbation $\mathcal{R}$ is found in the exponent of the three-metric $h_{ij}=a^2(t)e^{2 \mathcal{R}} \delta_{ij}$ in the space-time element $\mathrm{d}s^2 = - N^2 \mathrm{d}t^2 + h_{ij} (\mathrm{d}x^{i}+N^{i} \mathrm{d} t)(\mathrm{d}x^{j} + N^{j} \mathrm{d} t)$ where $N$ and $N^{i}$ are the lapse and shift, respectively.

In simple inflationary models the predicted PPS is a power law
\begin{align}
		\mathcal{P}(k) = A \left(\frac{k}{k_{\ast}}\right)^{n_s-1}
\end{align}
where $A$ is the amplitude measured at the wave number $k_{\ast}$ (the pivot scale) and $n_s$ is the spectral index, close to $1$.
Furthermore, non-Gaussianity is strongly suppressed so the $2$-point function provides a sufficient description of the statistics. The amplitude and spectral index can be related to the inflationary (slow roll) parameters. The amplitude is proportional to $(H/M_{\mathrm{Pl}})^2$ and inversely proportional to $\epsilon$. Reflecting the evolution of the scalar field during inflation the spectral index $n_s$ depends on $\eta$ also such that $n_s -1 = 2\eta - 4\epsilon$.
%where $\eta = -\ddot{\phi}/(H \dot{\phi}) \approx $. %The pivot scale is typically $k=0.05 \, \mathrm{Mpc}^{-1}$
%An attractive aspect of inflation is that it also quantitatively accounts for the presence of small inhomogeneities in the hot plasma needed to be present to grow by gravitational instability.
%Departures from slow-roll inflation generally lead to modifications of the power-law form of the PPS, imprinting characteristic scales, features, in the spectrum.
%Beyond slow-roll. The phenomenology is rich with examples.
%A more complicated scenario than single-field slow-roll inflation is expected 

%We shall be concerned with primordial power spectra with features.
There are inflationary scenarios where the resulting PPS differs from the simple power-law form and many possible realisations have been explored. For instance, this can be done by introducing a more complicated potential \cite{Hodges:1989dw}, multiple fields \cite{Salopek:1988qh}, inflation in stages \cite{Gottlober:1990um}~\cite{Adams:1997de}, violations of slow-roll \cite{Hodges:1990bf}, or a plateau in the potential that amplifies the power of fluctuations and allows primordial black hole production \cite{Ivanov:1994pa}.

After noting that the PPS need not be simple we now turn to how it is inferred from data. The PPS is related to the observed temperature fluctuations of the CMB in a linear way. If the temperature fluctuations $\Delta T(\hat{\mathbf{n}})$ are decomposed in spherical harmonics $\Delta T(\hat{\mathbf{n}}) = \sum a_{\ell m} Y_{\ell m}(\hat{\mathbf{n}})$ then the angular power spectrum $C_{\ell}$, which is the two-point function of the spherical harmonics $\langle a_{\ell m} a^{\ast}_{\ell' m'} \rangle = C_{\ell} \delta_{\ell \ell'} \delta_{m m'}$
can be related to the PPS by transfer functions $\Delta_\ell(k)$ such that
\begin{align}
		C_{\ell} = 4 \pi \int_{0}^{\infty} \mathrm{d}\log k \, \Delta_{\ell}^2(k) \mathcal{P}(k).
\end{align}
By discretising $\mathcal{P}(k) \to \mathbf{p}$ using a sufficiently fine grid the relation above can be written as a matrix equation $\mathbf{d} = \mathbf{W} \mathbf{p}$ where $C_{\ell}$ corresponds to $\mathbf{d}$. It is impossible to obtain the PPS $\mathbf{p}$ from the data by an inversion with $\mathbf{W}^{-1}$ as this matrix does not exist. Many different PPS can lead to the same data. However, it is possible to choose the most likely PPS given the data subject to the constraint that the PPS is not too `jagged', with this roughness parameterised by an integral of squared first derivatives of the PPS.

\begin{figure}
\begin{minipage}{0.45\linewidth}
\centerline{\includegraphics[width=1\linewidth]{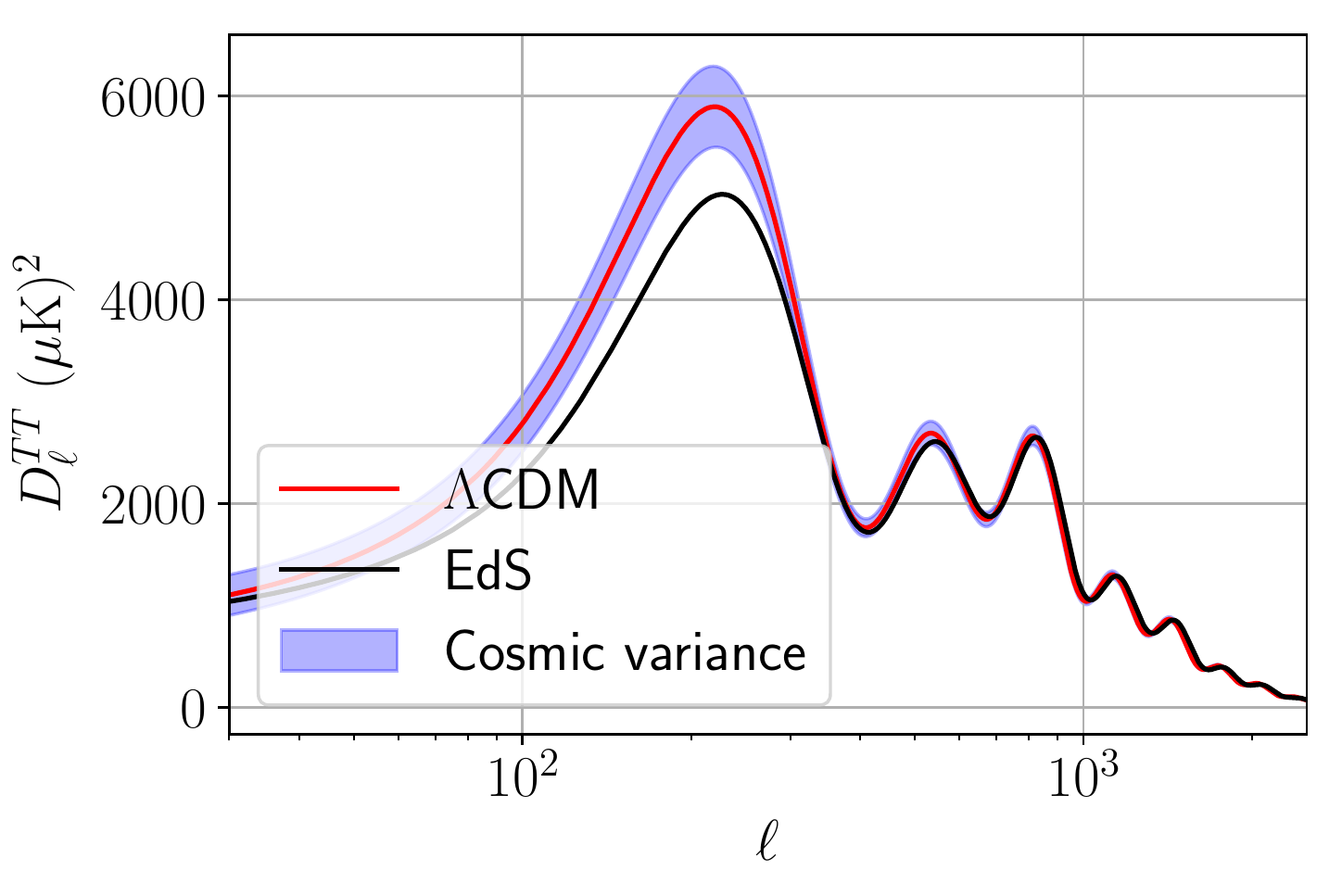}}
\end{minipage}
\hfill
\begin{minipage}{0.45\linewidth}
\centerline{\includegraphics[width=1\linewidth]{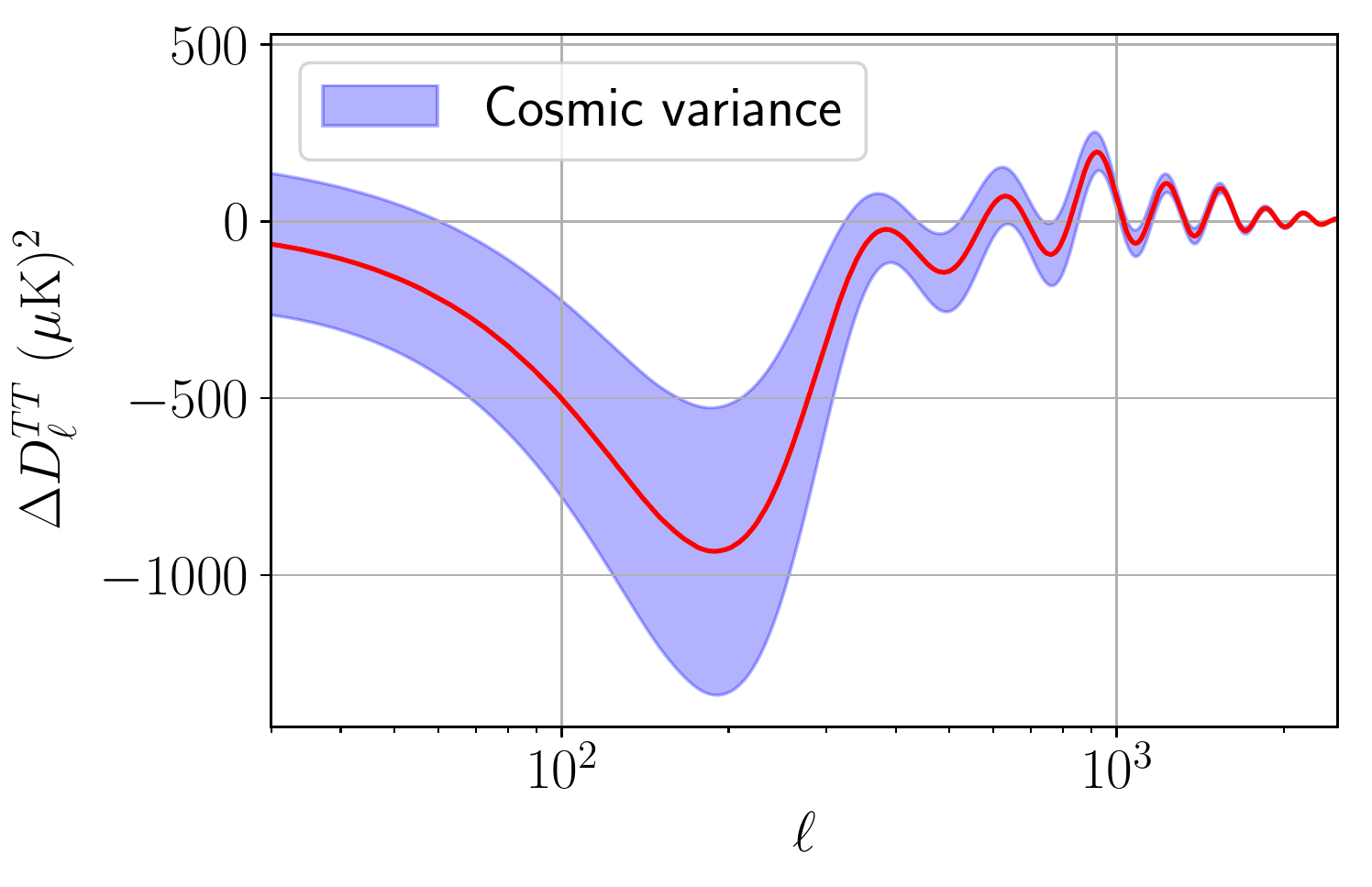}}
\end{minipage}
\caption[]{Left panel: The angular power spectrum $D_\ell \equiv \ell (\ell + 1) C_\ell / (2\pi)$ for the same power-law PPS assuming a no-$\Lambda$ cosmological model (black line), the best-fit $\Lambda$CDM cosmological model (red line) and its cosmic variance (purple band). Right panel: The difference between the angular power spectra (red line) compared with cosmic variance (purple band).}
\label{fig:ang}
\end{figure}

In addition, $\mathbf{W}$ is dependent on the choice of cosmological parameters as the growth of perturbations is sensitive to the background cosmology. Keeping the data $\mathbf{d}$ fixed it is possible to find a PPS $\mathbf{p}_1$ and a set of alternative cosmological parameters (having for instance $\Omega_\Lambda=0$) with the associated transfer function $\mathbf{W}_1$ which together produce the same data $\mathbf{d}$ as the current best-fit $\Lambda$CDM cosmological model with $\mathbf{W}_2$ and a power-law PPS $\mathbf{p}_2$. Then $\mathbf{d} = \mathbf{W}_2 \mathbf{p}_2$ and $\mathbf{d} = \mathbf{W}_1 \mathbf{p}_1$. It is clear that a power-law PPS will not work for a $\Omega_\Lambda=0$ model. This is illustrated in Figure~\ref{fig:ang} where the best-fit $\Lambda$CDM cosmology with a power-law PPS is compared to a no-$\Lambda$ \footnote{In this case, the EdS model, or more accurately, the CHDM model inspired by \cite{Hunt:2004vt} was considered.} model with a power-law PPS. The difference between the two spectra is visible beyond cosmic variance. By compensating with the addition and subtraction of power on particular scales, through oscillations, a physically plausible PPS may provide a good fit for an $\Omega_\Lambda=0$ model.
%\cite{Salopek:1988qh}
%Though varying with scale, the power-law PPS is otherwise without a characteristic scale. Departures from slow-roll inflation generally lead to modifications of the power-law form of the PPS. Characteristic scales, or features may be imprinted on it due to a more complicated inflationary trajectory. The changing background is directly reflected in the resulting PPS with early times corresponding to large scales, or low wave numbers. The phenomenology is rich. Fast roll followed by slow roll suppresses the PPS on large scales. The inflaton may encounter a complicated potential arising from its interactions with other fields. This can be a sharp change in the potential that imprints oscillations in the spectrum. Through non-canonical kinetic terms, the fluctuations may have their speed of propagation reduced, also affecting the power. As small $\epsilon$ leads to large fluctuations having a phase where $\epsilon$ is tiny in the form due to a potential with a plateau has been proposed as a mechanism for primordial black hole production \cite{Ivanov:1994pa}.

The many possible realisations of inflation lead us to look for an effective description which can encompass many of them as different histories of its parameters. The EFT of inflation has already been formulated \cite{Cheung:2007st}. The idea is that a part of the diffeomorphism invariance of general relativity, a space-dependent time shift, can be used to cancel the fluctuations in the scalar field. As the metric also transforms under this shift it now gets a complicated spatial dependence. The possible theories that respect the residual spatial symmetries are the remaining invariants which involve the $00$ metric component and the extrinsic curvature. %After imposing the desired background, taking into account constraints
It can be shown \cite{Chluba:2015bqa} that the second-order action of the EFT of inflation can be written in terms of the curvature perturbation $\mathcal{R}$ as
\begin{align}
		S_2 = M_{\mathrm{Pl}}^2 \int \mathrm{d}^3x \int \mathrm{d}\tau \, \epsilon(\tau) \, a^2(\tau) \left( (\mathcal{R}')^2/c_s^{2}(\tau) - (\partial_i \mathcal{R} )^2 \right) \label{eq:s2}
\end{align}
where $\epsilon(\tau)$ is the slow-roll parameter and $c_s(\tau)$ is the speed of sound \footnote{Matter sector couplings to the shift vector have been omitted which would contribute with one more term.}. A complicated inflationary history is now parameterised in the functions $\epsilon(\tau)$ and $c_s(\tau)$.

%\section{The EFT of the curvature perturbation}
%\section{Without $\mathbf \Lambda$}
\section{Establishing the dictionary}
Consider the second-order action of the EFT of inflation \eqref{eq:s2} and take the scale factor to be that of pure de Sitter space: $a(\tau) = -(H \tau)^{-1}$.

As previously noted, the PPS is the two-point function of the Fourier coefficients of the curvature perturbation $\mathcal{P}(k) \propto \langle \mathcal{R}_\mathbf{k}(\tau) \mathcal{R}_{\mathbf{k}'}(\tau) \rangle$. This expectation value is understood as that in quantum field theory $\langle 0| \hat{\mathcal{R}}_\mathbf{k}(\tau) \hat{\mathcal{R}}_{\mathbf{k}'}(\tau) |0 \rangle$ and it is to be evaluated at the end of inflation $\tau = 0$. The solution for the mode functions $\mathcal{R}_k$ and their derivative is known
\begin{align}
		\mathcal{R}_k(\tau) = \frac{iH(1+i k\tau) \exp(-ik \tau)}{2 M_{\mathrm{Pl}}\sqrt{\epsilon k^3}}, \qquad \mathcal{R}'_k(\tau) = \frac{iHk^2 \tau \exp(-ik \tau)}{2 M_{\mathrm{Pl}} \sqrt{\epsilon k^3 }} \label{eq:mf}.
\end{align}
when the EFT parameters do not vary.
The theory is quantised by $[\hat{a}_{\mathbf{k}},\hat{a}_{\mathbf{k}'}^{\dagger}] = (2\pi)^3 \delta^{(3)}(\mathbf{k}+\mathbf{k}') $
where
\begin{align}
		\mathcal{R}(\tau) = \int \frac{\mathrm{d}^3 k}{(2 \pi)^3} \left( \mathcal{R}_k(\tau) \hat{a}_{\mathbf{k}} \exp(i \mathbf{k} \cdot \mathbf{x}) + \mathcal{R}^{\ast}_{k}(\tau) \hat{a}_{\mathbf{k}}^{\dagger} \exp(-i \mathbf{k} \cdot \mathbf{x})  \right)
\end{align}
and the vacuum satisfies $\hat{a}_\mathbf{k} |0  \rangle = 0$.

The case of interest, the time-dependent slow-roll parameter is split into a constant and a varying part $\epsilon(\tau) = \epsilon + \Delta \epsilon(\tau) = \epsilon ( 1 + \Delta \epsilon/\epsilon(\tau))$. This turns the action into a solvable piece and an interacting piece with two terms:
\begin{align}
		S_2 = \epsilon M_{\mathrm{Pl}}^2 \int \mathrm{d}^3 x \int \mathrm{d}\tau a^2(\tau) \left( (\mathcal{R}')^2 - (\partial_i \mathcal{R})^2 \right) + M_{\mathrm{Pl}}^2 \int \mathrm{d}^3 x \int \, \mathrm{d}\tau \, \Delta \epsilon(\tau) a^2(\tau) ((\mathcal{R}')^2 - (\partial_i \mathcal{R} )^2 ).
\end{align}
Here the second integral is the interaction component with Lagrangian density
\begin{align}
		\mathcal{L}_{\mathrm{int}} = M_\mathrm{Pl}^2 \Delta \epsilon(\tau) a^2(\tau) ((\mathcal{R}')^2 - (\partial_i \mathcal{R})^2 ).
\end{align}

In the case of the speed of sound the split is made so that $1/c_s^2(\tau) \equiv 1 - (1 - 1/c_s^2(\tau)) = 1- u(\tau)$, defining $u(\tau) = 1- 1/c_s^2(\tau)$ as, here, excursions from $c_s = 1$ will be considered. The action becomes
\begin{align}
		S_2 = \epsilon M_{\mathrm{Pl}}^2 \int \mathrm{d}^3 x \int \mathrm{d}\tau a^2(\tau) \left( (\mathcal{R}')^2 - (\partial_i \mathcal{R})^2 \right) + \epsilon M_{\mathrm{Pl}}^2 \int \mathrm{d}^3 x \int \, \mathrm{d}\tau \, u(\tau) a^2(\tau) (\mathcal{R}')^2
\end{align}
identifying $\mathcal{L}_\mathrm{int} = \epsilon M^2_{\mathrm{Pl}} u(\tau) a^2(\tau) (\mathcal{R}')^2$. The corresponding Hamiltonian densities are the negatives of the Lagrangian densities $\mathcal{H}_\mathrm{int} = - \mathcal{L}_\mathrm{int}$.

Corrections to the PPS are corrections to the expectation value $\langle \mathcal{R}_\mathbf{k} \mathcal{R}_{\mathbf{k}'} \rangle$ due to the interacting parts. The fractional change in the PPS will turn out to be proportional to the fractional change in the EFT parameters which are typically $\sim 10\%$ and this justifies a perturbative treatment.

The expectation value $\langle \mathcal{R}_\mathbf{k} \mathcal{R}_{\mathbf{k}'} \rangle$ in the interacting theory is given by \cite{Weinberg:2005vy}
\begin{align}
		\langle \mathcal{R}_\mathbf{k}(\tau) \mathcal{R}_{\mathbf{k}'}(\tau) \rangle &= \sum_{n=0} i^{n} \int_{-\infty}^{\tau} \mathrm{d}\tau_n \int_{-\infty}^{\tau_n} \mathrm{d}\tau_{n-1} \cdots \nonumber \\ &\qquad \int_{-\infty}^{\tau_{2}} \mathrm{d}\tau_1 \left\langle 0 \huge| [\mathcal{H}_{\mathrm{int}}(\tau_1),[\mathcal{H}_{\mathrm{int}}(\tau_2), \cdots ,[\mathcal{H}_{\mathrm{int}}(\tau_n),\mathcal{R}_{\mathbf{k}}(\tau) \mathcal{R}_\mathbf{k'}(\tau)]]]] \huge| 0 \right\rangle
\end{align}
and is evaluated for $\tau=0$.
These are multiple integrals of expectation values over nested commutators of the interaction Hamiltonian density with the operator that is being considered. The number of nested commutators corresponds to the order of the perturbative correction. All terms reduce to a product of creation and annihilation operators and the answer is given by all possible Wick contractions.

The correction to the PPS from slow-roll parameter excursions is found to be
\begin{align}
		\Delta_{1} \mathcal{P}/\mathcal{P}(k) = \int_{-\infty}^{0} \frac{\mathrm{d} \tau}{\tau} \frac{1}{k \tau} \Delta \epsilon / \epsilon(\tau) ((1- 2 (k \tau)^2 ) \sin(2k \tau) - 2k \tau \cos(2 k \tau))
\end{align}
where the subscript $1$ indicates that the computation is the first order result. This can be inverted to give the slow-roll parameters in terms of the PPS modulation
\begin{align}
		\Delta \epsilon/ \epsilon(\tau) = \frac{2}{\pi} \int_{0}^{\infty} \frac{\mathrm{d}k}{k} \Delta_1 \mathcal{P}/\mathcal{P}(k) (2 \sin^2(k \tau)/(k \tau) - \sin(2 k \tau)). \label{eq:dicteps}
\end{align}
%On account of its change only leading to an interaction that has one operator
Since a speed of sound change only leads to an interaction term with one operator it is found to have a simpler relation to the induced modulation
\begin{align}
		\Delta_{1} \mathcal{P}/\mathcal{P}(k) = -k \int_{-\infty}^{0} \mathrm{d}\tau \, u(\tau) \sin(2k \tau)
\end{align}
which inverts to
\begin{align}
u(\tau) \equiv c_s^{-2}(\tau)-1 =  \frac{4}{\pi} \int_{-\infty}^{0} \frac{\mathrm{d} k}{k} \Delta \mathcal{P}/\mathcal{P}(k) \sin(2k \tau).
\end{align}
It turns out that the second-order correction to the PPS modulation from changes in either EFT parameter equals the square of the respective first-order correction: $\Delta_2 \mathcal{P}/\mathcal{P}(k) = (\Delta_1 \mathcal{P}/\mathcal{P}(k))^2$. Then the full modulation of the PPS is the sum of the first-order correction and its square
\begin{align}
		\Delta \mathcal{P}/\mathcal{P}(k) = \Delta_1 \mathcal{\mathcal{P}}/\mathcal{P}(k) + \Delta_2 \mathcal{P}/\mathcal{P}(k) = \Delta_1 \mathcal{\mathcal{P}}/\mathcal{P}(k) + (\Delta_1 \mathcal{P}/\mathcal{P}(k))^2 \label{eq:secord}
\end{align}
and so it is possible to isolate either EFT parameter for a given correction $Y \equiv \Delta \mathcal{P}/\mathcal{P}(k)$ by noticing that the relation between the integral transform ($J$) of the EFT parameter ($X$) $Z \equiv J(X)$ is related to the given correction $Y$ by a quadratic equation: $Y=Z+Z^2$ whose one solution is $Z = (-1+\sqrt{1+4 Y})/2$. Knowing the forward $J$ and inverse relation $J^{-1}$ between $X$ and $Z$ it is found that $X = J^{-1}((-1+\sqrt{1+4 Y})/2)$. In other words, the first-order dictionary can be used even in the second-order case once the \emph{given} (to be inverted) PPS modulation $\Delta \mathcal{P}/\mathcal{P}$ is replaced by an \emph{effective} PPS
\begin{align}
		\Delta^{\mathrm{eff}}_1 \mathcal{P}/\mathcal{P}(k) = \sqrt{1/4 + \Delta \mathcal{P}/\mathcal{P}(k) } -1/2 %\frac{1+ \sqrt{4 \Delta \mathcal{P} / \mathcal{P}(k)}}{2}.%
\end{align}
so that \begin{align}
		\Delta \epsilon/ \epsilon(\tau) = \int_{0}^{\infty}\frac{\mathrm{d}k}{k} \Delta^{\mathrm{eff}}_1 \mathcal{P}/\mathcal{P}(k) (2 \sin^2(k \tau)/(k \tau) - \sin(2 k \tau)),
\end{align}
yielding an expression for $\Delta \epsilon/\epsilon$ which is expected to receive $\sim (\Delta \mathcal{P}/\mathcal{P})^3$ order corrections and higher. For $\sim 20\%$ modulations of the PPS this corresponds to neglected corrections to $\Delta \epsilon/\epsilon$ of only $\sim 1\%$.
\section{Numerical check: toy model}
\begin{figure}
\centering
\centerline{\includegraphics[width=0.45\linewidth]{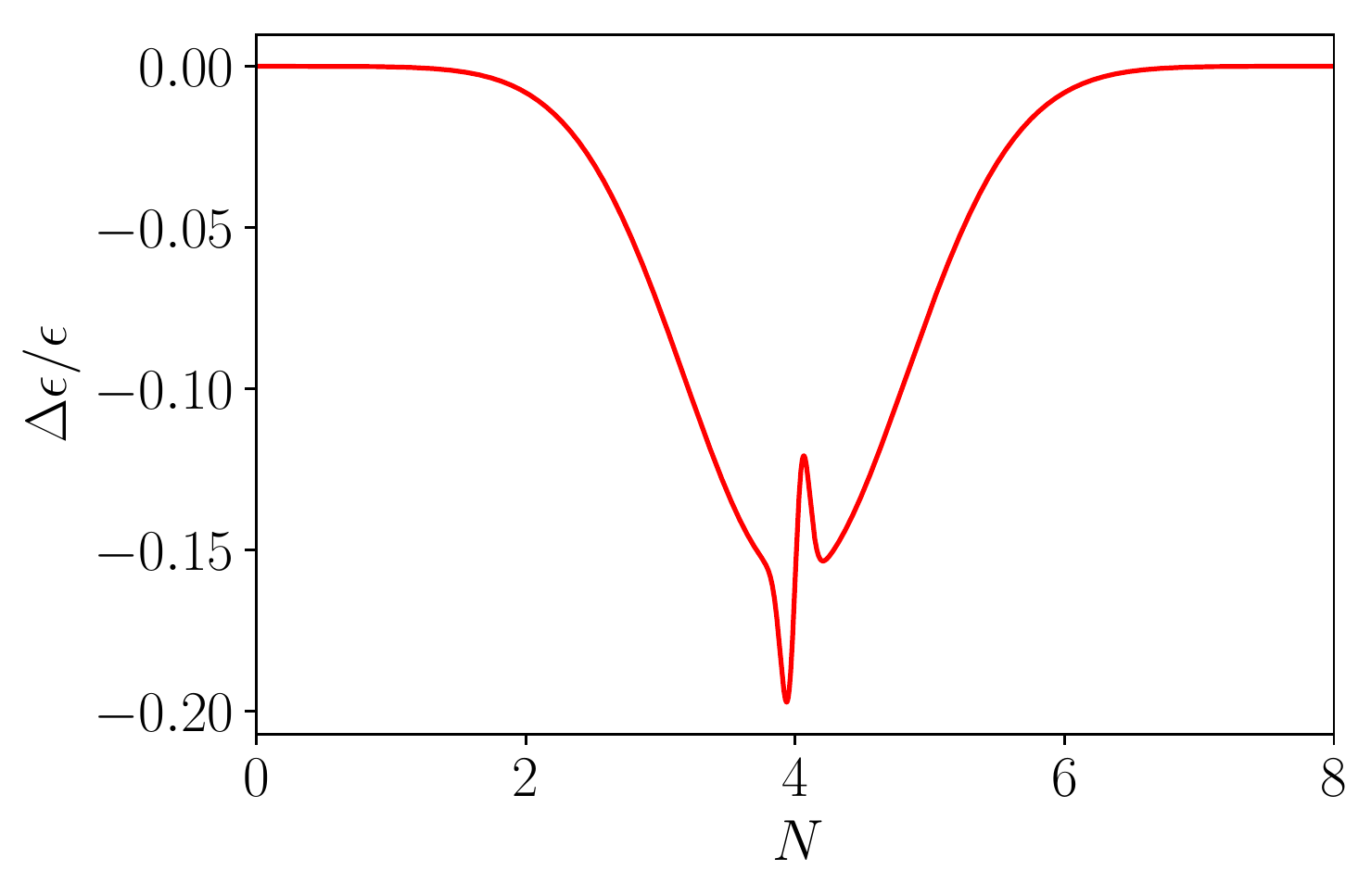}}
\caption[]{The toy model slow-roll parameter change \eqref{eq:toy} used for testing the dictionary.}
\label{fig:plottoy}
\end{figure}

The dictionary was tested against the numerical integration of the equation evolving the curvature perturbation mode by mode assuming a complicated change in the slow-roll parameter. The assumed fractional slow-roll parameter change shown in Figure~\ref{fig:plottoy} is that of a Gaussian and the derivative of a Gaussian
\begin{align}
		\Delta \epsilon / \epsilon(N) = c_1 e^{-(N-N_0)^2/\sigma_1^2} + c_2 (N-N_0) e^{-(N-N_0)^2/\sigma_2^2} \label{eq:toy}
\end{align}
with parameters $c_1=-0.159$, $c_2=0.99$, $\sigma_1=1.16$ and $\sigma_2=0.09$ centred around an $e$-fold value $N_0=4$ and constant $\epsilon = 10^{-4}$. The particular values are unimportant except that there is a slow component (first term) and a fast component (second term) to the variation set by $\sigma_1$ and $\sigma_2$, respectively. The choice of zero $e$-fold $N=0$ is conventional. It corresponds to the moment a given mode $k_0$ leaves the horizon.

\begin{figure}
\begin{minipage}{0.45\linewidth}
\centerline{\includegraphics[width=1\linewidth]{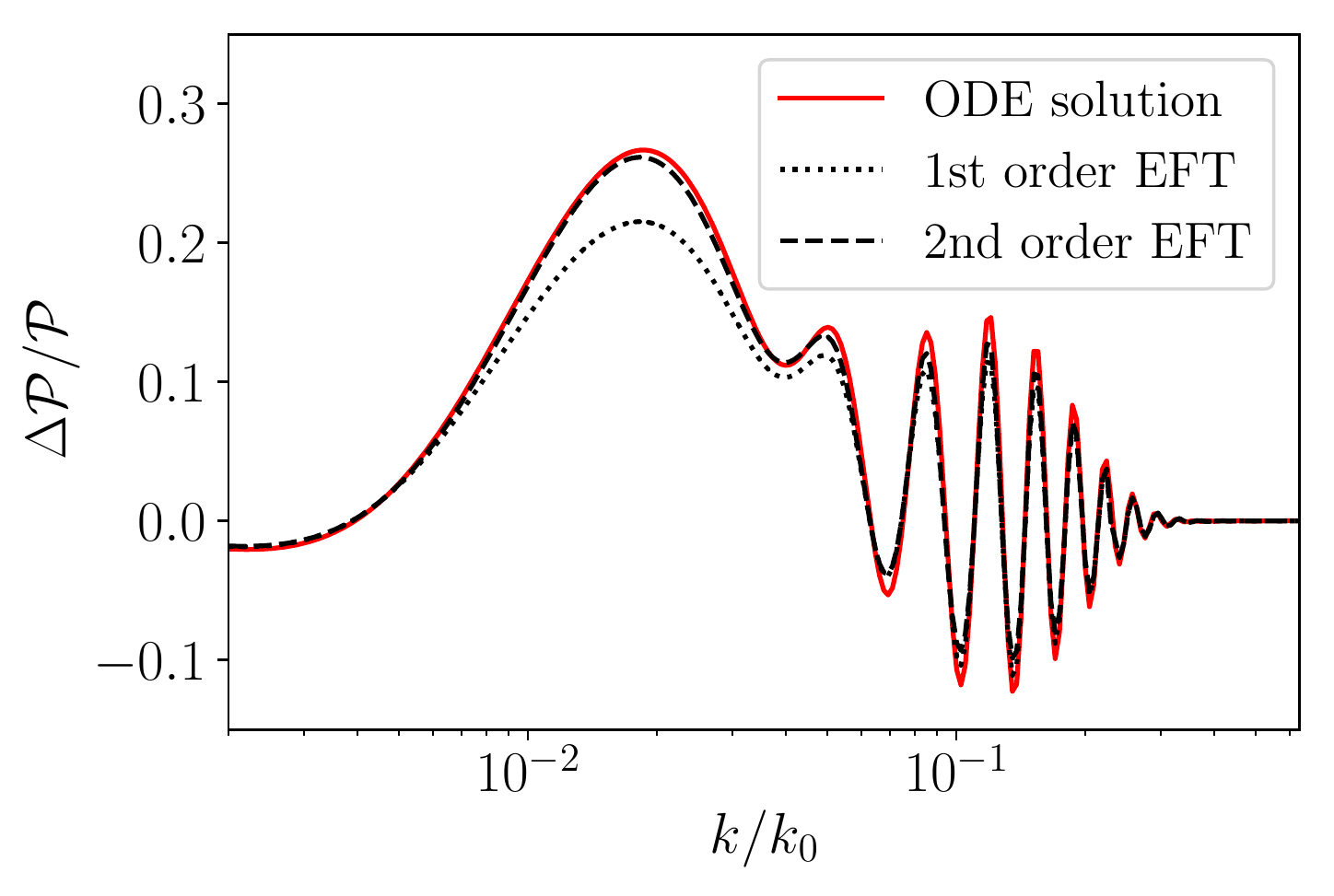}}
\caption[]{The fractional change in the PPS resulting from the slow-roll parameter change \eqref{eq:toy} solved numerically (red line), first-order dictionary (dotted black) and including second order contribution (dashed black).}
\label{fig:allthree}
\end{minipage}
\hfill
\begin{minipage}{0.45\linewidth}
\centerline{\includegraphics[width=1\linewidth]{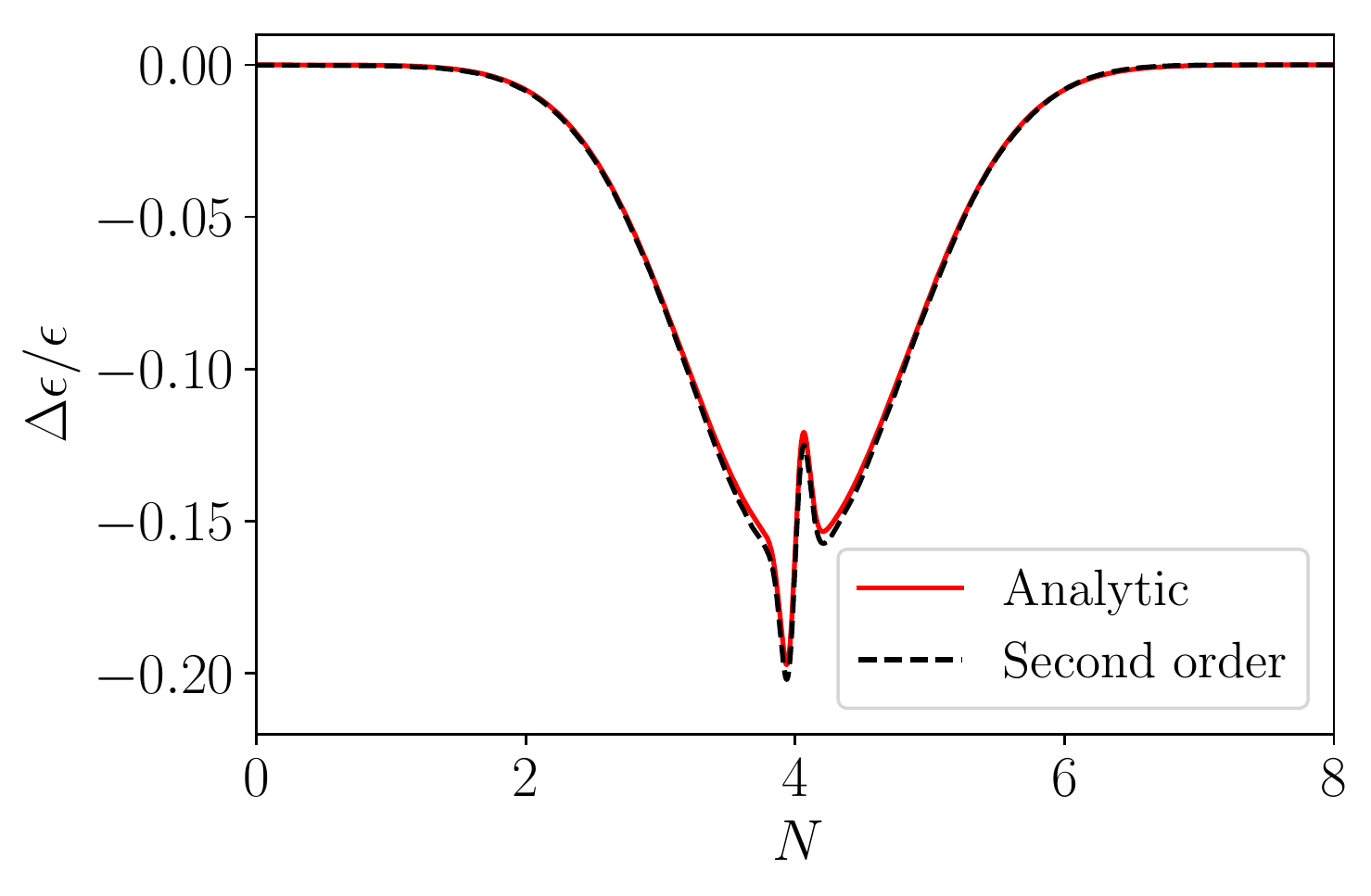}}
\caption[]{The known change in the slow-roll parameter (red line) and the estimated change (dashed black line) using the second-order corrected dictionary on the fractional PPS change (red line of Figure~\ref{fig:allthree}).}
\label{fig:invertsec}
\end{minipage}
%\caption[]{same figure with draft option (left), normal (center) and rotated (right)}

\end{figure}

Having specified $\Delta \epsilon/\epsilon$, the \emph{exact} fractional change in the PPS $\Delta \mathcal{P}/\mathcal{P}$ that the $\Delta \epsilon/\epsilon$ induces can be obtained by solving the equation for the curvature perturbation
\begin{align}
		\frac{\mathrm{d}^2\mathcal{R}_k}{\mathrm{d}N^2} + \left(3- \epsilon(N)+ \frac{\epsilon'(N)}{\epsilon(N)}\right) \frac{\mathrm{d}\mathcal{R}_k}{\mathrm{d}N} + \left(\frac{k}{aH}\right)^2 \mathcal{R}_k = 0
\end{align}
mode by mode and computing $k^3 |\mathcal{R}_k|^2 / (2 \pi^2)$, having imposed the initial conditions respecting \eqref{eq:mf}. This is to be compared with applying the first-order order formula for the modulation of the PPS \eqref{eq:dicteps} or obtaining an answer that takes second-order corrections into account by adding in addition a term equal to the square of the first-order result \eqref{eq:secord}. These three cases are shown in Figure~\ref{fig:allthree} where the numerical solution (full red line) is compared to the first order correction (dotted black line) and second order corrected result (dashed black line). The inclusion of the second-order contribution improves the dictionary to agree with the numerical result within a band $\sim (\Delta \mathcal{P}/\mathcal{P})^3$. As the fractional changes in the toy model are large, up to $25\%$, the first-order result was not expected to account for the whole change.

%From the slow-roll parameter change it is possible to reconstruct a scalar field potential that reproduces this behaviour. Even though the induced features are significant the potential can be quite benign. The constraint is on the derivative of the potential. It is shown in Figure~\ref{fig:pot}
% \begin{figure}
% \centering
% \centerline{\includegraphics[width=0.45\linewidth]{dVfi}}
% \caption[]{Derivative of reconstructed potential $\mathrm{d}V/\mathrm{d}\phi$ for $\epsilon$ change given by toy model \eqref{eq:toy} (orange line) and a constant $\epsilon$ for comparison (dashed blue line).}
% \label{fig:pot}
% \end{figure}

Having confirmed that the dictionary works it will now be applied to real data. %Though not shown here, the inclusion of the second-order contribution makes the calculated PPS modulation fall

\section{Making both $\mathbf \Lambda$ and no-$\mathbf \Lambda$ cases fit}
\begin{figure}
\begin{minipage}{0.45\linewidth}
\centerline{\includegraphics[width=1\linewidth]{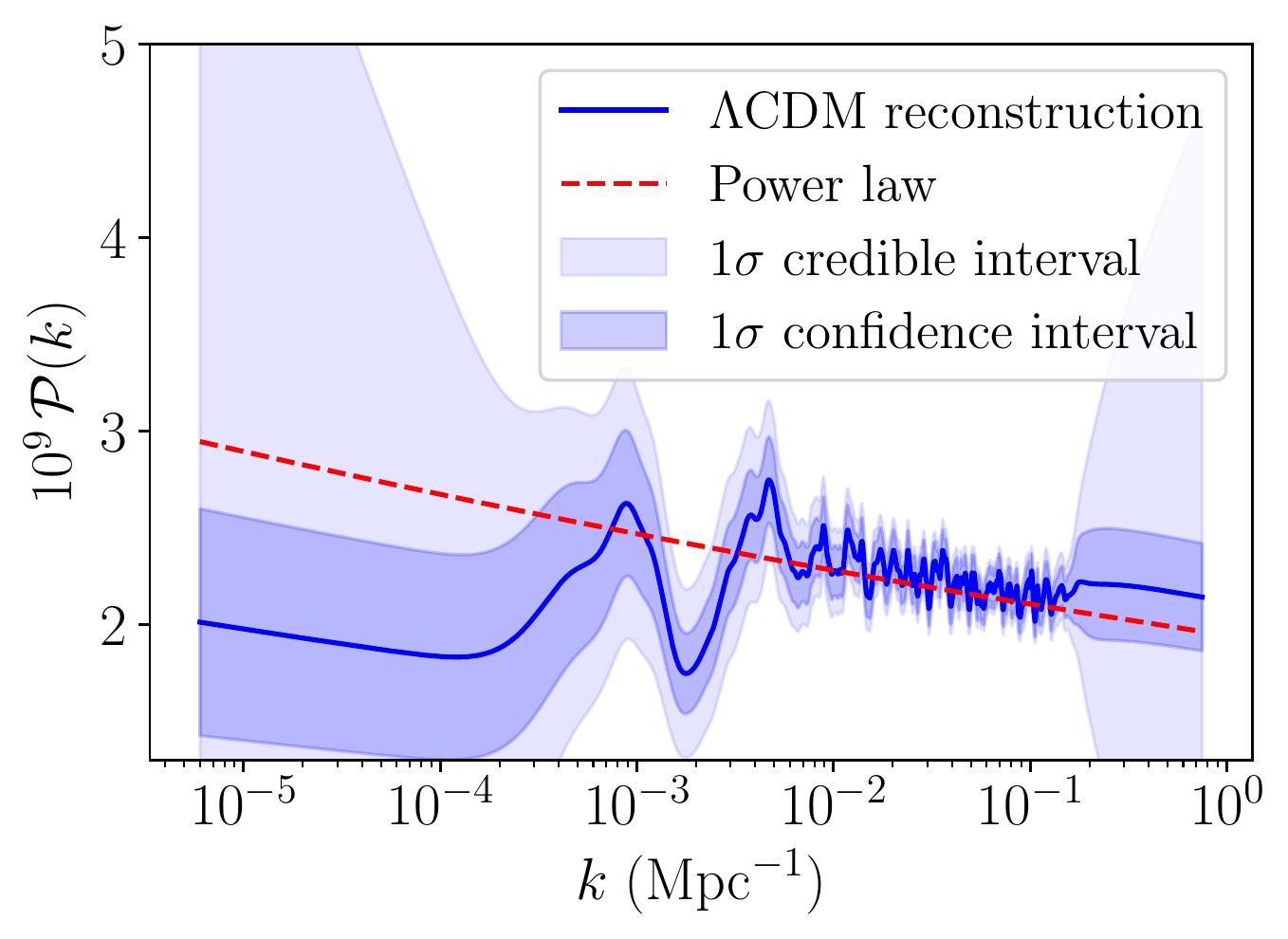}}
\end{minipage}
\hfill
\begin{minipage}{0.45\linewidth}
\centerline{\includegraphics[width=1\linewidth]{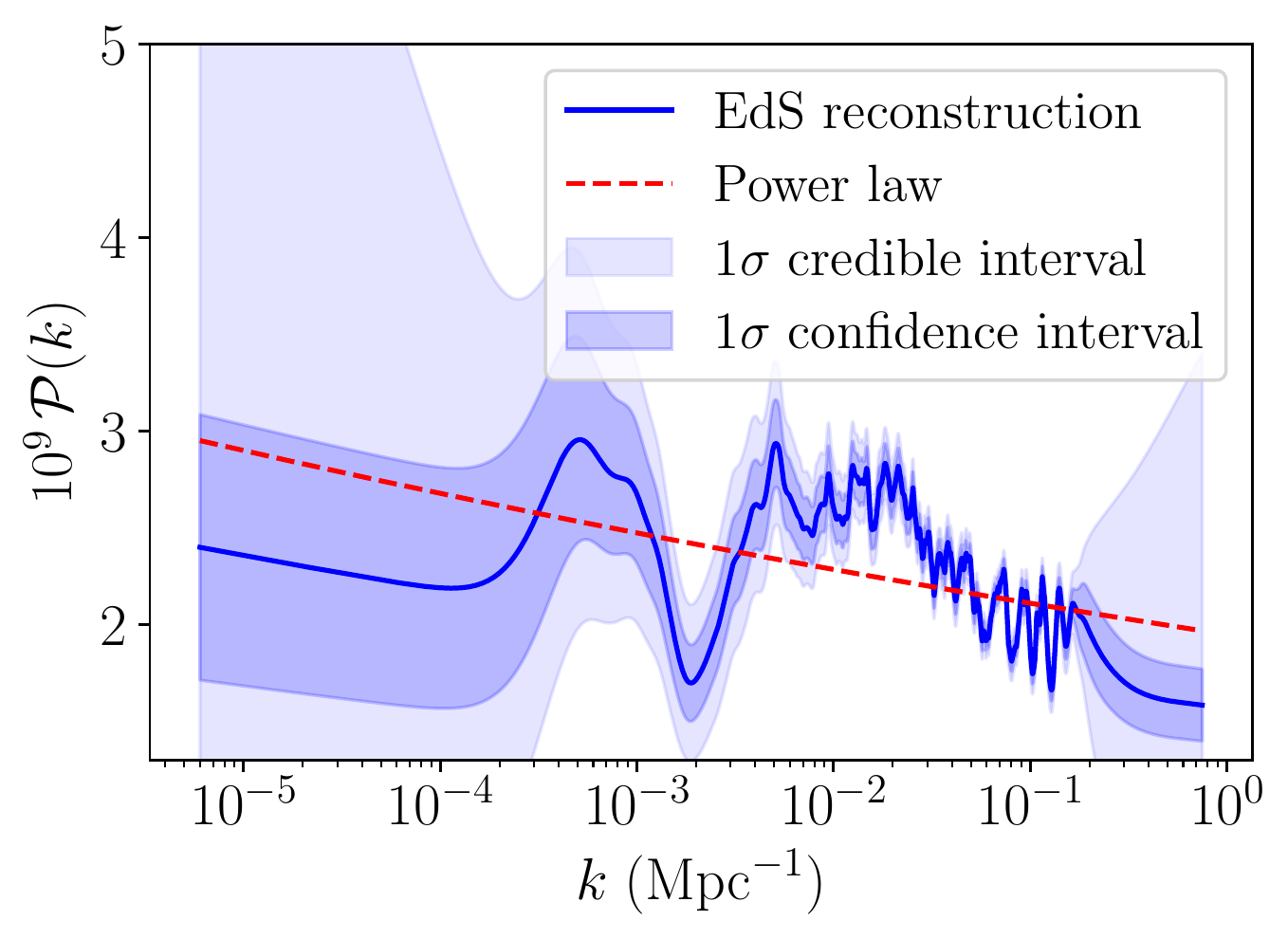}}
\end{minipage}
\caption[]{The PPS (full blue line) estimated from Planck temperature data assuming the best-fit $\Lambda$CDM model (left panel) and the no-$\Lambda$ model (right panel) both compared with a power law (dashed red line) including $1\sigma$ confidence (frequentist) and credible (Bayesian) intervals.}
\label{fig:ppsest}
\end{figure}

The PPS was estimated from Planck Release 2 (2015) \cite{Aghanim:2015xee} temperature ($TT$) data imposing a penalty on the square of the first derivative of the PPS, thus \emph{favouring} smooth spectra, but not excluding rough ones. The details of this procedure are found in \cite{Hunt:2015iua} and \cite{Hunt:2013bha}. As the estimation of the PPS is itself dependent on the assumed cosmological parameters there is one for each set adopted cosmological model. The estimated PPS assuming the best-fit $\Lambda$CDM model and the EdS model are seen in Figure~\ref{fig:ppsest} where both estimated PPS are compared to a power law.

In \emph{both} cases, the reconstructed PPS have lower power on scales near $2 \times 10^{-3} \, \mathrm{Mpc}^{-1}$ compared to the power law. This comes from low power in the $TT$ angular power spectrum near $\ell = 23$. While the PPS of $\Lambda$CDM seems otherwise consistent with a power-law PPS the EdS model requires a prominent feature and additional oscillations on smaller scales.

\begin{figure}
\begin{minipage}{0.45\linewidth}
\centerline{\includegraphics[width=1\linewidth]{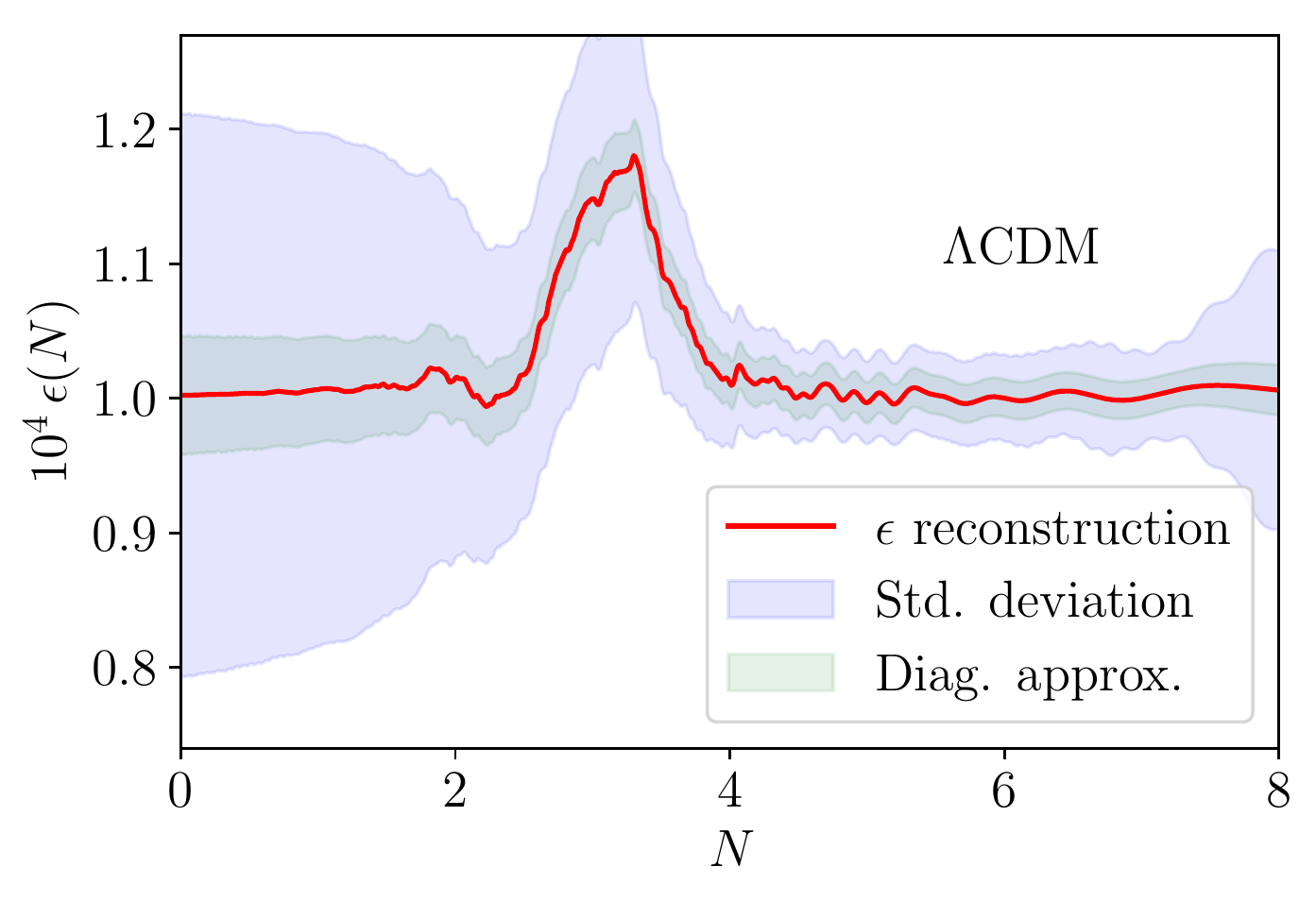}}
\end{minipage}
\hfill
\begin{minipage}{0.45\linewidth}
\centerline{\includegraphics[width=1\linewidth]{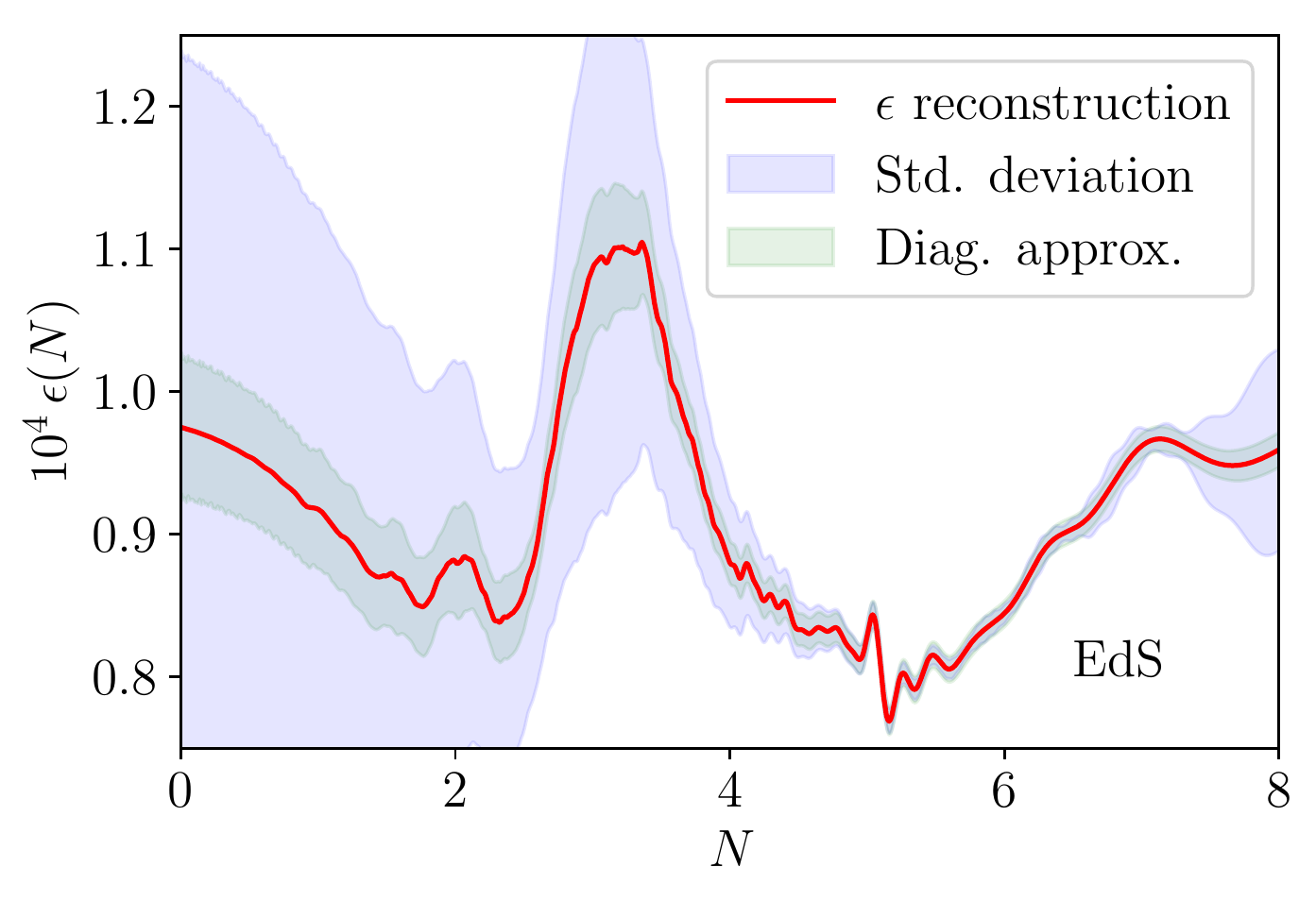}}
\end{minipage}
\caption[]{The slow roll-parameter changes (red line) reconstructed from the PPS estimates from Planck data assuming the best-fit $\Lambda$CDM cosmology (left panel) and the no-$\Lambda$ cosmological model (right panel) including $1\sigma$ uncertainty estimates (purple band). The green band is a simplified uncertainty estimate that results from assuming the covariance matrices of the PPS to be diagonal.}
\label{fig:epsest}
\end{figure}

This is reflected in the slow-roll parameter changes reconstructed from these PPS estimates using the dictionary with second-order corrections which are shown in Figure~\ref{fig:epsest}. Both reconstructions favour a peak near $N=5$ to explain the low-$\ell$ anomaly. In addition the no-$\Lambda$ case requires a trough of characteristic size $\Delta N \sim 1$ including a sharp feature ($\Delta N \sim 0.1$) in the middle, similar to the toy model.

These findings should be treated with caution as an inspection of the plots by eye does not reveal the correlations present between the data points. This is illustrated in the difference between the uncertainty in the estimated $\epsilon$ when using only the diagonal values of the covariance matrix associated with the PPS estimates (green band of Figure~\ref{fig:epsest}) and using the full covariance matrix (purple band). This continues to hold for the $\epsilon$ estimates which have correlations among neighbouring $e$-fold values which this plot does not show. A full statistical analysis returning maximum-likelihood estimates of the slow-roll parameter changes should be performed. This was beyond the scope of this work whose focus was the establishment of the dictionary. The statistical significance of the presence of features is left for future work.

\section{Conclusion}
Though features may be interesting from the point of view of elucidating the mechanism behind cosmological inflation they are currently not required for the best-fit $\Lambda$CDM model. For alternative cosmological models such as EdS, however, features are necessary for an acceptable fit to CMB data. As the angular power spectrum is sensitive to both the cosmological parameters and the PPS, in order that the angular power spectrum stay fixed under a change of cosmological parameters, the PPS must be modulated in a specific way to ensure this. In these cases, the necessary PPS modulation is very specific.

While the underlying mechanisms for the production of features can be diverse, the predictions are ultimately for the statistics of the adiabatic curvature perturbation. The EFT of inflation is the EFT of the curvature perturbation where to second order the curvature perturbations arising from a more complicated inflationary scenario can be parameterised by changes in the EFT parameters: the slow-roll parameter, which is also the order parameter of the theory, and the speed of sound.

A dictionary was found relating the modulations of the PPS $\Delta \mathcal P/ \mathcal P(k)$ to the parameters of the EFT, and vice versa. First, corrections to the PPS due to excursions in either the speed of sound $c_s$ or the slow-roll parameter $\epsilon$ were computed using perturbation theory. The resulting integral transforms relating the excursions to the PPS were then inverted to also express the excursions of the respective parameters in terms of a given modulation of the PPS. Both the forward relation, expressing the modulation of the PPS to the effective parameters, and its inverse relation were computed to second order in the effective parameters, leading to relations that are expected to be of the order $(\Delta \mathcal{P}/\mathcal{P})^3$. For modulations as high as $\sim 20\%$, this amounts to a less than $1\%$ error in the slow-roll/speed of sound parameter.

The dictionary was verified by checking that a numerical integration of the equation evolving the curvature perturbation $\mathcal{R}(k)$ from early to late times assuming a given excursion of the slow-roll parameter $\Delta \epsilon/\epsilon(\tau)$ would produce the same modulation of the PPS $\Delta \mathcal{P}/\mathcal{P}(k)$ as that given by the dictionary.

The dictionary was then applied to obtain the slow-roll parameter excursions $\Delta \epsilon/\epsilon(\tau)$ from estimates of the PPS $\Delta \mathcal{P}/\mathcal{P}(k)$ reconstructed from Planck Release 2 data. As the estimated PPS depends on the assumed cosmologial model both the case of the best-fit $\Lambda$CDM cosmological model and an alternative EdS cosmological model was considered. The slow-roll parameter changes reconstructed from Planck data suggest that sharp features are needed for a no-$\Lambda$ cosmology to fit Planck data. %By reconstructing the required scalar field potential simple realisations of single-field inflation that would produce such PPS modulations $\Delta \mathcal{P}(k)$ were also found.
\section{Outlook}
If the observed acoustic peaks have a primordial component arising from a more complicated inflationary theory, this interacting theory is expected to have \emph{additional} signatures manifesting as primordial non-Gaussianity. The EFT provides the tools \cite{Achucarro:2012fd}, the cubic action and perturbation theory, to compute the resulting form of non-Gaussianity. However, preliminary studies indicate that this is currently too weak to be observed in the CMB. In this way, the exercise of finding the PPS subject to the constraint that it provides a good fit for an alternative cosmological model goes beyond finding a fit and has phenomenological consequences. In this particular case, the specificity of the modulation is benecifial because it gives a fixed non-Gaussian template to search for.
%Confronting alternatives with the combination of large scale structure data and CMB data is an ongoing.

A full statistical analysis of the reconstructed slow-roll parameter changes should be performed to elucidate exactly which reconstructed slow-roll parameter excursions are needed to fit the no-$\Lambda$ cosmology.

Here, estimates of the PPS have been used to reconstruct the EFT parameters. That is not strictly necessary. It is possible to go directly from the cosmological data set to the EFT parameters, circumventing the PPS. As the relation between the EFT parameters and the PPS are now known, and the relation between the PPS and the angular power spectrum is also known, the composition of the two maps, or a matrix multiplication in the case of two linear relations, provides the transfer function from EFT parameters to the cosmological data set. The inversion, going from the cosmological data set to the EFT parameters, can then be accomplished under a smoothness constraint on the EFT parameters.

A more pertinent and ongoing investigation is confronting the no-$\Lambda$ model with the combination of large scale structure data and CMB data. Large scale structure surveys peg the matter power spectrum at multiple redshifts where the only difference between the matter power spectra at the given redshifts is the cosmological evolution, depending only on the cosmological parameters of the theory.
\section*{Acknowledgments}
Many thanks to the organisers for giving me the opportunity to present this work. Special thanks to Jacques Dumarchez and Yannick Giraud-H\'{e}raud. Thanks to director Jean Tr\^{a}n Thanh V\^{a}n, his wife Kim and the local organisers at ICISE for hosting this wonderful conference. Thanks to Peter Hansen at the Discovery Center for supporting my trip. Thanks to Subir Sarkar, Paul Hunt and Subodh Patil for the collaboration. Thanks also to Subir, Paul, Sebastian von Hausegger and Sunny Vagnozzi for comments on the draft.


\section*{References}

\begin{thebibliography}{99}

%\cite{Durakovic:2019kqq}
\bibitem{Durakovic:2019kqq}
  A.~Durakovic, P.~Hunt, S.~P.~Patil and S.~Sarkar,
  %``Reconstructing the EFT of Inflation from Cosmological Data,''
  SciPost Phys.\  {\bf 7} (2019) 049
  %doi:10.21468/SciPostPhys.7.4.049
  [arXiv:1904.00991 [astro-ph.CO]].
  %%CITATION = doi:10.21468/SciPostPhys.7.4.049;%%

%\cite{Cheung:2007st}
\bibitem{Cheung:2007st}
  C.~Cheung, P.~Creminelli, A.~L.~Fitzpatrick, J.~Kaplan and L.~Senatore,
  %``The Effective Field Theory of Inflation,''
  JHEP {\bf 0803} (2008) 014
  %doi:10.1088/1126-6708/2008/03/014
  [arXiv:0709.0293 [hep-th]].
  %%CITATION = doi:10.1088/1126-6708/2008/03/014;%%
  %719 citations counted in INSPIRE as of 03 Nov 2019

%\cite{Chluba:2015bqa}
\bibitem{Chluba:2015bqa}
  J.~Chluba, J.~Hamann and S.~P.~Patil,
  %``Features and New Physical Scales in Primordial Observables: Theory and Observation,''
  Int.\ J.\ Mod.\ Phys.\ D {\bf 24} (2015) no.10,  1530023
  %doi:10.1142/S0218271815300232
  [arXiv:1505.01834 [astro-ph.CO]].
  %%CITATION = doi:10.1142/S0218271815300232;%%
  %96 citations counted in INSPIRE as of 29 Oct 2019

%\cite{Achucarro:2012fd}
\bibitem{Achucarro:2012fd}
  A.~Ach\'{u}carro, J.~O.~Gong, G.~A.~Palma and S.~P.~Patil,
  %``Correlating features in the primordial spectra,''
  Phys.\ Rev.\ D {\bf 87} (2013) no.12,  121301
  %doi:10.1103/PhysRevD.87.121301
  [arXiv:1211.5619 [astro-ph.CO]].
  %%CITATION = doi:10.1103/PhysRevD.87.121301;%%
  %84 citations counted in INSPIRE as of 30 Oct 2019

%\cite{Weinberg:2005vy}
\bibitem{Weinberg:2005vy}
  S.~Weinberg,
  %``Quantum contributions to cosmological correlations,''
  Phys.\ Rev.\ D {\bf 72} (2005) 043514
  %doi:10.1103/PhysRevD.72.043514
  [hep-th/0506236].
  %%CITATION = doi:10.1103/PhysRevD.72.043514;%%
  %608 citations counted in INSPIRE as of 30 Oct 2019

%\cite{Hunt:2015iua}
\bibitem{Hunt:2015iua}
  P.~Hunt and S.~Sarkar,
  %``Search for features in the spectrum of primordial perturbations using Planck and other datasets,''
  JCAP {\bf 1512} (2015) 052
  %doi:10.1088/1475-7516/2015/12/052
  [arXiv:1510.03338 [astro-ph.CO]].
  %%CITATION = doi:10.1088/1475-7516/2015/12/052;%%
  %24 citations counted in INSPIRE as of 01 Nov 2019
%\cite{Hunt:2013bha}
\bibitem{Hunt:2013bha}
  P.~Hunt and S.~Sarkar,
  %``Reconstruction of the primordial power spectrum of curvature perturbations using multiple data sets,''
  JCAP {\bf 1401} (2014) 025
  %doi:10.1088/1475-7516/2014/01/025
  [arXiv:1308.2317 [astro-ph.CO]].
  %%CITATION = doi:10.1088/1475-7516/2014/01/025;%%
  %48 citations counted in INSPIRE as of 01 Nov 2019

 %\cite{Hunt:2004vt}
 \bibitem{Hunt:2004vt}
   P.~Hunt and S.~Sarkar,
   %``Multiple inflation and the WMAP 'glitches',''
   Phys.\ Rev.\ D {\bf 70} (2004) 103518
   %doi:10.1103/PhysRevD.70.103518
   [astro-ph/0408138].
   %%CITATION = doi:10.1103/PhysRevD.70.103518;%%
   %107 citations counted in INSPIRE as of 01 Nov 2019

%\cite{Ivanov:1994pa}
\bibitem{Ivanov:1994pa}
  P.~Ivanov, P.~Naselsky and I.~Novikov,
  %``Inflation and primordial black holes as dark matter,''
  Phys.\ Rev.\ D {\bf 50} (1994) 7173.
  %doi:10.1103/PhysRevD.50.7173
  %%CITATION = doi:10.1103/PhysRevD.50.7173;%%
  %154 citations counted in INSPIRE as of 02 Nov 2019

%\cite{Salopek:1988qh}
\bibitem{Salopek:1988qh}
  D.~S.~Salopek, J.~R.~Bond and J.~M.~Bardeen,
  %``Designing Density Fluctuation Spectra in Inflation,''
  Phys.\ Rev.\ D {\bf 40} (1989) 1753.
  %doi:10.1103/PhysRevD.40.1753
  %%CITATION = doi:10.1103/PhysRevD.40.1753;%%
  %721 citations counted in INSPIRE as of 02 Nov 2019

%\cite{Hodges:1989dw}
\bibitem{Hodges:1989dw}
  H.~M.~Hodges, G.~R.~Blumenthal, L.~A.~Kofman and J.~R.~Primack,
  %``Nonstandard Primordial Fluctuations From a Polynomial Inflaton Potential,''
  Nucl.\ Phys.\ B {\bf 335} (1990) 197.
  %doi:10.1016/0550-3213(90)90177-F
  %%CITATION = doi:10.1016/0550-3213(90)90177-F;%%
  %92 citations counted in INSPIRE as of 02 Nov 2019

%\cite{Hodges:1990bf}
\bibitem{Hodges:1990bf}
  H.~M.~Hodges and G.~R.~Blumenthal,
  %``Arbitrariness of inflationary fluctuation spectra,''
  Phys.\ Rev.\ D {\bf 42} (1990) 3329.
  %doi:10.1103/PhysRevD.42.3329
  %%CITATION = doi:10.1103/PhysRevD.42.3329;%%
  %93 citations counted in INSPIRE as of 02 Nov 2019

%\cite{Adams:1997de}
\bibitem{Adams:1997de}
  J.~A.~Adams, G.~G.~Ross and S.~Sarkar,
  %``Multiple inflation,''
  Nucl.\ Phys.\ B {\bf 503} (1997) 405
  %doi:10.1016/S0550-3213(97)00431-8
  [hep-ph/9704286].
  %%CITATION = doi:10.1016/S0550-3213(97)00431-8;%%
  %184 citations counted in INSPIRE as of 02 Nov 2019

%\cite{Aghanim:2015xee}
\bibitem{Aghanim:2015xee}
  N.~Aghanim {\it et al.} [Planck Collaboration],
  %``Planck 2015 results. XI. CMB power spectra, likelihoods, and robustness of parameters,''
  Astron.\ Astrophys.\  {\bf 594} (2016) A11
  %doi:10.1051/0004-6361/201526926
  [arXiv:1507.02704 [astro-ph.CO]].
  %%CITATION = doi:10.1051/0004-6361/201526926;%%
  %673 citations counted in INSPIRE as of 02 Nov 2019

%\cite{Gottlober:1990um}
\bibitem{Gottlober:1990um}
  S.~Gottlober, V.~Muller and A.~A.~Starobinsky,
  %``Analysis of inflation driven by a scalar field and a curvature squared term,''
  Phys.\ Rev.\ D {\bf 43} (1991) 2510.
  %doi:10.1103/PhysRevD.43.2510
  %%CITATION = doi:10.1103/PhysRevD.43.2510;%%
  %78 citations counted in INSPIRE as of 02 Nov 2019

\end{thebibliography}
\end{document}